\documentstyle[preprint,eqsecnum,aps,epsf]{revtex}
\newif\iftightenlines\tightenlinesfalse
\tightenlines\tightenlinestrue
\begin{document}
%
\def\pT{p_T^{\phantom{7}}}
\def\MW{M_W^{\phantom{7}}}
\def\ET{E_T^{\phantom{7}}}
\def\bh{\bar h}
\def\lm{\,{\rm lm}}
\def\lo{\lambda_1}                                              
\def\lt{\lambda_2}
\def\pslt{p\llap/_T}
\def\eslt{E\llap/_T}
\def\to{\rightarrow}
\def\Re{{\cal R \mskip-4mu \lower.1ex \hbox{\it e}}\,}
\def\Im{{\cal I \mskip-5mu \lower.1ex \hbox{\it m}}\,}
\def\SU{SU(2)$\times$U(1)$_Y$}
\def\te{\tilde e}
\def\tl{\tilde l}
\def\ttau{\tilde \tau}
\def\tg{\tilde g}
\def\tga{\tilde \gamma}
\def\tnu{\tilde\nu}
\def\tell{\tilde\ell}
\def\tq{\tilde q}
\def\tw{\widetilde W}
\def\tz{\widetilde Z}
%
%
\preprint{\vbox{\baselineskip=14pt%
   \rightline{FSU-HEP-931104}\break 
   \rightline{SSCL-Preprint-531}\break
   \rightline{UH-511-777-93}
}}
\title{DETECTING SLEPTONS AT HADRON COLLIDERS\\
AND SUPERCOLLIDERS}
\author{Howard Baer$^1$, Chih-hao Chen$^1$, Frank Paige$^2$\\
 and Xerxes Tata$^3$}
\address{
$^1$Department of Physics,
Florida State University,
Tallahassee, FL 32306 USA
}
\address{
\hfill $^2$Superconducting Supercollider Laboratory,
Dallas, TX 75237 USA, and \hfill\break
Brookhaven National Laboratory, Upton, NY 11973 USA
}
\address{
$^3$Department of Physics and Astronomy,
University of Hawaii,
Honolulu, HI 96822 USA
}
\date{\today}
\maketitle
\begin{abstract}
We study the prospects for detecting the sleptons of the Minimal 
Supersymmetric Standard Model at hadron colliders and supercolliders.
We use ISAJET 7.03 to simulate charged slepton and sneutrino
pair production, incorporating slepton and sneutrino cascade decays
into our analysis. We find that 
even with an accumulation of $\sim 1$fb$^{-1}$ of integrated luminosity,
it will be very difficult to detect sleptons beyond the reach of LEP
at the Fermilab Tevatron $p\bar{p}$ collider, due to a large background
from $W$ pair production.
At LHC, sleptons of mass up to 300 GeV ought to be detectable via
dilepton signal as long as it is possible to veto central jets with
$p_T \geq 25$ GeV with high efficiency.

\end{abstract}
\medskip
\pacs{PACS numbers: 14.80.Ly, 11.30.Pb, 13.85.Qk}
%
%
%
\section{Introduction}

It has long been known that supersymmetrization of the Standard Model (SM)
leads to the stabilization of fundamental scalar masses, provided that
super-partner masses are less than $\sim 1$ TeV\cite{MSSM}. 
More recently, it has
been observed that the precision measurements of gauge couplings at 
scale $M_Z$ by the LEP experiments are also consistent with the simplest
supersymmetric grand unified model (but not with minimal SU(5)) 
if sparticle masses are
$\sim 1$ TeV\cite{SUGRA}. These motivations have prompted the recent
re-examination of signatures for supersymmetry (SUSY) 
at various hadron colliders.
Attention has been focussed mainly on the strongly interacting
sparticles, squarks and gluinos\cite{BKT}, 
charginos and neutralinos\cite{INOS}, and Higgs bosons of SUSY\cite{HIGGS}. 

In contrast, SUSY signals from slepton production at hadron colliders 
have received rather
limited attention, probably due to the smallness of the cross sections.
Early studies\cite{BENT} were limited to slepton production via decays
of real $W$ and $Z$ bosons at the CERN ${\rm Sp\bar{p}S}$ Collider. 
Del Aguila and Ametller\cite{AGUILA} have performed the
most detailed study of these signals at hadron supercolliders and concluded
that sleptons should be detectable with masses up to 250~GeV at
the Large Hadron Collider (LHC) and up to 350~GeV at the unfortunately
now defunct Superconducting Supercollider (SSC).
To the best of our knowledge,
there does not exist any analysis of slepton signals for the experiments at 
the Tevatron which are expected to accumulate a combined integrated
luminosity of 100~pb${}^{-1}$ by the end of the current Tevatron Run~1B.

The purpose of this paper is to
reexamine the prospects for detecting sleptons and sneutrinos
at both the Fermilab Tevatron Collider
and at LHC. We have
improved on the analysis of Ref.\cite{AGUILA} in several respects.
Unlike these authors who assume that sleptons always directly decay
to the LSP, we have incorporated all their cascade decay patterns.
Generally speaking, these decays lead to reduction of the acollinear dilepton
signal and are particularly important for the left-type sleptons. Also,
we have used ISAJET 7.03 for our analysis, and so have a somewhat
more realistic simulation of the signal and backgrounds as compared to previous
analyses which were done at the parton level. This is especially important for
the simulation of the extent of the hadronic activity in the signal events
which, as we will see, will be useful in discriminating the signal from top
quark background. 

We work within the framework of the Minimal Supersymmetric Standard 
Model (MSSM)\cite{MSSM}, which is the simplest supersymmetric
extension of the SM. 
In the MSSM, for each generation of leptons $(\nu_{\ell} ,\ell )$, there exists
two spin zero charged sleptons, $\tell_L$ and $\tell_R$, and a neutral
spin zero particle, the sneutrino $\tnu_L$. 
The mixing between charged sleptons
is proportional to the corresponding {\it fermion} mass, and so,
is completely negligible except possibly in the tau sector. 
Since we will mainly be concerned with the first two families
of sleptons, we assume the L and R weak eigenstates
are also the physical particles for the remainder of this paper.

Various experiments have already put lower bounds on slepton and sneutrino
masses. The ASP Collaboration\cite{ASP}
which searched for single photon events produced via the reaction
$e^+e^-\to\tz_1\tz_1\gamma$
has excluded selectrons with mass $m_{\te}<58$~GeV if L and R selectrons
are degenerate, and $\tz_1$ is a massless photino. This bound is
sensitive to the composition of the LSP\cite{DKT}, and further, disappears
if $m_{\tz_1}>12.5$~GeV. Non-observation of
acoplanar dilepton plus missing energy events 
in the four LEP experiments\cite{LEP} leads to the bounds, 
\begin{eqnarray}
m_{\tell_L} > 45\ {\rm GeV},\nonumber \\
m_{\tell_R} > 45\ {\rm GeV},
\eqnum{1}
\end{eqnarray}
which are valid unless $m_{\tz_1}\simeq m_{\tell}$.
Constraints can also be placed on sneutrino masses at LEP, even though
sneutrinos are expected to decay invisibly\cite{BDT,LEP}. The combined LEP
experiments require non-standard contributions to the $Z$ boson 
invisible width to
be $\Delta\Gamma_{\rm inv} <17$~MeV at 95\% CL, which implies\cite{LEP}
%
%
\begin{eqnarray}
& m_{\tnu_L} > 37\ {\rm GeV}\ ({\rm single\ sneutrino}),\nonumber
& \\
& m_{\tnu_L} > 41.7\ {\rm GeV}\ ({\rm three\ degenerate\
sneutrinos}). &
\eqnum{2}
\end{eqnarray}

Our choices for sparticle masses and mixing angle parameters
are guided by the framework of supergravity grand unified theories.
In supergravity models, supersymmetry breaking leads to a common
mass for sfermions at the unification scale. The degeneracy of sfermions 
present at the unification scale
is broken when these masses are evolved down to the weak scale. 
Since the lepton masses are neglible, the slepton
masses can then be written as\cite{MASSES},
\begin{eqnarray}
m_{\tell_L}^2 &=& m_{\tilde q}^2-0.73 m_{\tilde g}^2-0.27 M_Z^2\cos 2\beta ,
\eqnum{3a}\\
m_{\tell_R}^2 &=& m_{\tilde q}^2-0.78 m_{\tilde g}^2-0.23 M_Z^2\cos 2\beta ,
\eqnum{3b}\\
m_{\tilde\nu_L}^2 &=& m_{\tilde q}^2-0.73 m_{\tilde g}^2+0.5 M_Z^2\cos 2\beta ,
\eqnum{3c}
\end{eqnarray}
where $m_{\tilde q}^2$ is the squark mass squared averaged 
over all the flavours. 

We see that for $m_{\tilde{q}} \gg m_{\tilde{g}}$,
the squarks are basically degenerate with the sleptons; significant
splitting between the masses of the three slepton types is possible
only when squarks and gluinos are rougly degenerate, in which case,
sleptons are considerably lighter than the squarks.
For the convenience of the reader,
we have plotted in Fig.~1 the various slepton masses as a function of 
$m_{\tilde{g}}$
where we take the average squark masses to be {\it a}) $m_{\tq}=m_{\tg}$,
and {\it b}) $m_{\tq}=2m_{\tg}$. We show results for $\tan\beta =2$ and 20.
Thus, from Fig.~1a, we see that if squark and gluino masses are nearly equal,
there is an expected large splitting of the various slepton masses, with the
sneutrino being the lightest slepton for low gluino masses, while $\tell_R$
is lightest for gluino masses larger than about 300 GeV. 
Furthermore, there is a large splitting between squark and slepton masses,
with sleptons being considerably lighter than squarks. In this situation,
the leptonic decays of neutralinos, and sometimes, also of
charginos can be significantly enhanced\cite{INOS}. Finally, we see
from Fig.~1b, that if all the sleptons and the squarks (other than $t$-squarks
which do not concern us here) are significantly heavier than gluinos, these
must be essentially degenerate.

The rest of this paper is organized as follows. In Sec.~2, we discuss
slepton production and decay at hadron colliders, and discuss
details of our simulation. In Sec.~3, we investigate slepton signals
and backgrounds relevant for Tevatron experiments, while in Sec.~4, we 
examine slepton signals and relevant backgrounds at the LHC. We conclude
in Sec.~5 with a brief summary of our results.

\section{Slepton production, decay and simulation}

At hadron colliders, sleptons are dominantly pair-produced
via Drell-Yan (DY) process mediated by
a (virtual) $W$, $\gamma$ or $Z$ in the $s$-channel. The rate for
slepton production
via $WW$\cite{GHM} or $gg$ fusion\cite{AGUILA} 
processes is smaller by at least an
order of magnitude at the LHC and is entirely negligible
at Tevatron energies.
Sleptons can also be singly produced 
via decays of charginos or neutralinos produced by the cascade decays
of squarks and gluinos. Since such events will be difficult to sort 
out from other cascade
decay patterns of gluinos and squarks\cite{BKT,BTW}, we will
focus on direct slepton production via the 
DY mechanism in this paper.

The cross section for the production of charged slepton-sneutrino pairs 
is given by
\begin{eqnarray}
{{d\sigma } \over {d\hat{t}}} (d\bar{u}\to W\to \tell_L \bar{\tnu}_L )
={{g^4 |D_W (\hat{s} )|^2}\over {192\pi \hat{s}^2}}
(\hat{t}\hat{u}-m_{\tell_L}^2 m_{\tnu_L}^2 ).
\eqnum{4}
\end{eqnarray}
For $\tilde{\ell_L}$ pair production, 
the cross section is given by\cite{DEQ}
\begin{eqnarray}
\lefteqn{ {{d\sigma}\over {d\hat{t}}} (q\bar{q}\to\gamma^* ,
Z\to \tell_L\bar{\tell}_L )
={{e^4}\over {24\pi \hat{s}^2}}
(\hat{t}\hat{u}-m_{\tell_L}^4 ) }\nonumber \\
& & \quad \times \left\{ {{q_{\ell}^2 q_q^2}\over {\hat{s}^2}}+
 ( \alpha_{\ell}-\beta_{\ell} )^2 (\alpha_q^2 +\beta_q^2 )|D_Z (\hat{s} )|^2 +
 {{2 q_{\ell} q_q \alpha_q (\alpha_{\ell}-\beta_{\ell} ) 
(\hat{s}-M_Z^2)}\over {\hat{s}}}
|D_Z (\hat{s})|^2 \right\},
\eqnum{5}
\end{eqnarray}
where $D_V(\hat{s})={1 / ({\hat{s}-M_V^2+iM_V\Gamma_V})}$ and the
$\alpha$'s, $\beta$'s, and charge assignments $q$ are given in
Ref.~\cite{TATA}. The cross section for sneutrino pair
production can be obtained by replacing $\alpha_\ell$, $\beta_\ell$,
$q_{\ell}$ and 
$m_{\tilde{\ell}}$ by $\alpha_{\nu}$, $\beta_{\nu}$, 0 and $m_{\tilde{\nu}}$,
respectively, whereas
for $\ell_R$ pair production
one substitutes $\alpha_{\ell}-\beta_{\ell}\to \alpha_{\ell}+\beta_{\ell}$,
and $m_{\tell_L}\to m_{\tell_R}$ in Eq.(5).

The left-sleptons dominantly decay via gauge interactions into charginos or 
neutralinos via the (kinematically accessible) two body decays,
\begin{eqnarray}
\tell_L &\to& \ell\ +\tz_i ,\nonumber \\
\tell_L &\to& \nu_{\ell}\ +\tw_j .
\eqnum{6}
\end{eqnarray}
and
\begin{eqnarray}
\tnu_L &\to& \nu_{\ell} \ +\tz_i ,\nonumber \\
\tnu_L &\to& \ell \ +\tw_j .
\eqnum{7}
\end{eqnarray}

If the sleptons are relatively light,
only the decay to the LSP may be possible, so that a light sneutrino
decays invisibly.
Heavier sleptons, can also decay via the chargino or other neutralino 
modes. Unless suppressed by phase space these decays are important and,
because they proceed via the larger SU(2) gauge coupling, frequently
dominate the direct decay to the LSP.  The daughter charginos and neutralinos further
decay until the cascade terminates in the stable LSP ($\tz_1$).

In contrast, the SU(2) singlet charged sleptons $\tell_R$ only decay
via their U(1) gauge interactions, so that in the limit of vanishing lepton
Yukawa coupling their decays to charginos are forbidden. These thus decay
via,
\begin{eqnarray}
\tell_R\to \ell\ +\tz_i ,
\eqnum{8}
\end{eqnarray}
Frequently, the branching
fraction for the $\tell_R\to \ell\tz_1$ mode is large even for rather
high values 
of $m_{\tell_R}$; these decays are, therefore, a potential source of
very hard isolated leptons at hadron colliders.
The dependence of the branching fractions for the various decays of sleptons 
and sneutrinos on the MSSM parameters has been studied in Ref.~\cite{BBKMT}
to which we refer the reader.

The slepton production processes and decay modes discussed above 
have all been incorporated into
the simulation program ISAJET 7.03\cite{ISAJET}. Briefly, for a given input
parameter set 
$( m_{\tg}, m_{\tq},m_{\tell_L},m_{\tell_R},m_{\tnu_L},\mu ,\tan\beta ,m_A )$,
the routine ISASUSY calculates all sparticle masses and branching fractions
to various decay modes. ISAJET then produces slepton pairs 
according to probabilities given by the above production formulae
convoluted with EHLQ Set~1 structure functions\cite{EHLQ}. These sleptons then
decay via the various cascades with appropriate
branching fractions as given by the MSSM. Radiation of 
initial and final state
partons is also included in ISAJET. Final state quarks and gluons are 
hadronized, and
unstable particles are decayed until stable final states are reached.
Underlying event activity is also modeled in our simulation.

\section{Sleptons at the Fermilab Tevatron collider}

The total DY production cross sections for slepton
pairs via $p\bar{p}$ collisions at Tevatron collider 
energy $\sqrt{s}=1.8$~TeV is shown in Fig.~2 as a function of slepton mass. 
On account of the LEP constraints, slepton production can only occur
via off-shell $W$ or $Z$ exchanges, and so,
each individual production cross section is typically 
below the 1 pb level. However, summing over the four processes shown,
as well as 
summing over the $e$ and $\mu$ families, 
can push the total slepton cross section
above the pb level.

The dominant production cross section comes from $W^*\to\tell\bar{\tnu_L}$. 
Unless the sneutrino is heavier than the chargino or the $\widetilde{Z_2}$,
it will decay invisibly via $\tnu\to\nu\tz_1$. Since the dominant
decays of the selectron are $\tell_L\to\ell\tz_1$, or if allowed, 
$\tell_L\to\tw_1\nu_{\ell}$, the $W^*$ production mode will result in 
single-hard-isolated-lepton
plus missing energy events. Such event topologies have large backgrounds due to
direct $W\to \ell\nu$ and $W\to\tau\to\ell$ decays, where $\ell$ can be
either $e$ or $\mu$. Similarly, $Z^*\to\tnu_L\bar{\tnu}_L$ will lead usually
to little observable activity in the final state, and is not a promising 
search mode if sneutrino decays to the chargino are forbidden. 
The modes $\gamma^*,Z^*\to \tell_L\bar{\tell}_L$ or 
$\tell_R\bar{\tell}_R$ followed by each slepton decaying via 
$\tell\to \ell\tz_1$ can lead to acollinear, hard isolated dilepton plus
missing energy events with little jet activity, 
which constitutes the most promising signature
for Tevatron collider searches. Backgrounds to the dilepton signature
come from $W$ pair production, from $t\bar t$ production, and from
$Z\to\tau\bar{\tau}$ production. It should be remembered that the 
decay $\tnu_L\to \tw_1 \ell$ rapidly dominates sneutrino decays
when it is kinematically allowed; in this case sneutrino production
is an additional source of acollinear dilepton pairs at the Tevatron.

In addition, there could be other
SUSY processes which mimic this dilepton signature, for instance,
DY production of chargino pairs, $\gamma^*,Z^*\to\tw_1\overline{\tw}_1$,
followed by $\tw_1\to \ell\nu_{\ell}\tz_1$ decays. Dilepton events can
also come from gluino and squark pair production\cite{BKT}, but these
events should be accompanied by substantial jet activity.

To assess slepton detection prospects at the Tevatron collider, we 
simulated two cases of slepton production: we take
%
%
\begin{description}
\item[\rm \ \ \ \ \ \ \ \ \ \ Case~1: ] $m_{\tg} =m_{\tq}=-\mu =150$ GeV,
$\tan\beta =2$ 
\item[\rm \ \ \ \ \ \ \ \  \ \ Case~2: ] $m_{\tg} =m_{\tq}=-\mu =200$ GeV,
$\tan\beta =2$ 
\end{description}
The above parameters are consistent with predictions from supergravity GUT
models with radiative electroweak symmetry breaking\cite{AN}. 
The corresponding slepton masses can be read off Fig.~1a,
and are listed in Table I as well. Case 1
leads to a somewhat lighter sparticle spectrum than case 2. In fact,
it is very close to the region of parameter space excluded by the CDF\cite{CDF}
experiment from their analysis of the $\eslt$ data sample. We have shown
it to illustrate the difficulty of detecting sleptons even when the
sparticles are relatively light.

We use the toy calorimeter simulation package ISAPLT to model detector effects.
We simulate calorimetry with cell size 
$\Delta\eta\times\Delta\phi =0.1\times 0.1$, which extends between
$-4<\eta <4$ in pseudorapidity. We take hadronic (electromagnetic) 
energy resolution to be $70\% /\sqrt{E_T}$ ($15\% /\sqrt{E_T}$). 
Jets are coalesced
within cones of $R=\sqrt{\Delta\eta^2 +\Delta\phi^2} =0.7$ using
the ISAJET routine GETJET. Clusters with $E_T>15$ GeV
are labelled as jets.
Muons and electrons are classified as isolated if they have $p_T>10$ GeV,
$|\eta (\ell )|<3$,
and the visible activity within a cone of $R =0.4$ about the lepton 
direction is less than $E_T({\rm cone})=5$ GeV.

We then impose the following cuts designed to select signal events, while
vetoing SM backgrounds from W pair, $\tau\bar{\tau}$ and top quark pair 
production:
\begin{itemize}
\item require {\it two} isolated same flavor leptons with 
$p_T(\ell )>15$ GeV,
\item require missing transverse energy $\eslt >20$ GeV,
\item require number of jets $n({\rm jets})=0$,
\item require transverse opening angle 
$30^o <\Delta\phi (\ell\bar{\ell} )<150^o$.
\end{itemize}

Cross sections after these cuts are listed in Table I in femtobarns.
The dominant SM background with $m_t =150$~GeV comes from $WW$
production. No events from 
$t\bar t$ or $\tau\bar{\tau}$ production passed our cuts, yielding upper
limits on background from these sources. The $t \bar t$ background
estimate may be somewhat optimistic since we have assumed an
idealized calorimeter covering $\vert\eta\vert < 4$ to determine the
rejection from the jet veto.

After the collection of 1 fb$^{-1}$ of data, we see that Case 1 would yield
a cross-section at the $3\sigma$ level above $WW$ background, while
Case 2 is only $1.5\sigma$ above background. Of course, related
processes such as chargino pair production also yield signal events, and
could be factored in either as signal or background. Chargino pair
rates for case 1 and 2 are listed in Table I as well for comparison.
In Case 1, approximately $2/3$ of the slepton signal events come from
$\tell_R\bar{\tell_R}$ production, while the other $\sim 1/3$ comes
from $\tell_L\bar{\tell_L}$; sneutrinos make a negligible contribution.
However, for Case 2, dileptons come nearly equally from $RR$ and $LL$
production, although now the $LL$ component contains a substantial
contribution from sneutrino pair production.

Since $WW$ production 
leads to an equal number of $e\mu$ 
events (using the above cuts), it might be possible to compare the rate
for $e\bar{e}$ and $\mu\bar{\mu}$ production to the rate for 
$e\bar{\mu} +\bar{e}\mu$ production.
Evaluating $R={{N(e\bar{e}+\mu\bar{\mu})} / {N(e\bar{\mu}+\bar{e}\mu )}}$
for integrated luminosity of 1 fb$^{-1}$ yields in Case 1,
$R=1.4\pm 0.3$, and in Case 2, $R=1.25\pm 0.25$, {\it i.e.} only a one
standard deviation effect. Incorporation of more realistic
detector effects and efficiencies would surely reduce these rates, leading us
to conclude that detection of a slepton signal at the Tevatron collider would 
be extremely difficult. 

\section{Slepton Search at the LHC}

The higher energy and higher luminosity available at the LHC
will considerably enhance slepton production rates relative to
the Tevatron, and leads to the possibility of detecting sleptons 
beyond the reach of LEP 200.
 In Fig.~3, we plot the slepton and sneutrino pair production
cross sections at $\sqrt{s}= 14$ TeV (LHC), once again
using EHLQ Set 1 structure functions. 
For a design luminosity of $3\times 10^{4}$ pb$^{-1}$/yr,
we see that at LHC, for instance, $pp\to \ell_R\bar{\ell}_R X$ can result in
about 2400 (12) events annually for $m_{\tell_R}=100$ (400)~GeV.
Summing over
L and R slepton types, flavors and generations considerably enhances these 
rates. 

For masses in the 100-400 GeV range, the cascade decays of sleptons
are very
important, and can lead to final states containing many isolated leptons
and jets. We consider first the same-flavor isolated dilepton 
signature, then turn to the assessment of single lepton and 
other multi-lepton signatures.

To examine the possibility of detecting sleptons at the LHC, we examine
four cases:
%
%
\begin{description}
\item[\rm \ \ \ \ \ \ \ \ \ \ Case~3: ] $m_{\tg} =m_{\tq}=-\mu =200$
GeV, $\tan\beta =2$ 
\item[\rm \ \ \ \ \ \ \ \ \ \ Case~4: ] $m_{\tg} =m_{\tq}=-\mu =400$
GeV, $\tan\beta =2$ 
\item[\rm \ \ \ \ \ \ \ \ \ \ Case~5: ] $m_{\tg} =m_{\tq}=-\mu =600$
GeV, $\tan\beta =2$ 
\item[\rm \ \ \ \ \ \ \ \ \ \ Case~6: ] $m_{\tg} =m_{\tq}=-\mu =800$
GeV, $\tan\beta =2$, 
\end{description}
for which slepton masses are $\sim 100,\ 200,\ 300$ and 400 GeV,
respectively. Exact slepton and sneutrino masses may be read off Fig.~1a.

For LHC, we again use the toy calorimeter 
simulation package ISAPLT.
We simulate calorimetry with cell size 
$\Delta\eta\times\Delta\phi =0.05\times 0.05$, which extends between
$-5.5<\eta <5.5$. We take hadronic 
energy resolution to be $50\% /\sqrt{E_T}$ for $|\eta |<3$, and to be
a constant 
$10\%$ for $3<|\eta |<5.5$, to model the effective $p_T$ resolution of
the forward calorimeter including the effects of shower spreading.

We take electromagnetic resolution to be $15\% /\sqrt{E_T}$. 
Jets are coalesced
within cones of $R=\sqrt{\Delta\eta^2 +\Delta\phi^2} =0.7$ using
the ISAJET routine GETJET. For the purpose of jet veto (essential to 
eliminate top quark background),
clusters with $E_T>25$ GeV
are labelled as jets.
Muons and electrons are classified as isolated if they have $p_T>20$ GeV,
$|\eta (\ell )|<2.5$,
and the visible activity within a cone of $R =0.3$ about the lepton 
direction is less than $E_T({\rm cone})=5$ GeV.

\subsection{Dilepton signature at the LHC}

For each of the four cases above, we generated 5000 slepton and sneutrino 
events in the ratio expected in the MSSM,
and examined backgrounds from $WW$ (50K events) and $t\bar t$ with
$m_t =150$ GeV (1.3M events).  The background events have been forced to 
have primary leptonic decays. Unlike at the Tevatron, the major
background at the LHC comes from $t\bar t$ production, and has a cross 
section of about 1500 pb for our choice of top mass.

To detect hard, isolated dilepton events, we impose the following cuts:
\begin{itemize}
\item A. require {\it two} isolated same flavor leptons, each with 
$p_T(\ell )>20$ GeV,
\item B. require missing transverse energy $\eslt >100$ GeV,
\item C. require number of jets with $|\eta |<3 $ to be $n(jets)=0$
(central jet veto),
\item D. require transverse opening angle 
$\Delta\phi (\vec{p_T}(\ell\bar{\ell}') ,\vec{\eslt} )>160^o$ ,
\end{itemize}
and
\begin{itemize}
\item E. make a slepton mass dependent cut on $p_T(\ell )>p_{T_c}$, and
$\Delta\phi (\ell\bar{\ell}' )<\phi_c$, which we optimize (as discussed
below) depending on the slepton mass.
\end{itemize}

The cross section after each cut is listed in Table II, along with the
per cent of cross section that was cut.
Cut A is more severe for background than for signal in part because
it selects out the leptonic branching fractions, which are smaller for
the background processes.
As might be anticipated,
the efficiency
of cut B ($\eslt$ cut) is sensitively dependent on slepton mass, 
since higher mass sleptons yield a harder $\eslt$ spectrum. Cut C, the central
jet veto, is very effective at cutting out $t\bar t$ events, since the
auxiliary $b$-jets are frequently hard and central. Cut D is an additional
cut designed to eliminate some fraction of $t\bar t$ events with soft $b$-jets;
these jets, which fail to pass the jet requirements, affect the direction
of the leptons and the $\eslt$ vector.

Finally, for cut E (high $p_{T}(\ell )$ and dilepton opening angle cut),
we examined a matrix of $p_{T_c}$ and $\phi_c$ values to find 
optimal signal to background levels. 
Results are given in Table III. We find that the choice ($p_{T_c}$, $\phi_c$)
= (40 GeV, $90^o$) [(80 GeV, $140^o$)] works well for the case of relatively
light [heavy] sleptons. For the transition region around $m_{\tell} = 200$ GeV,
we exhibit the results with both sets of cuts in Table III. When we found
zero events in our simulation, we have used the 1 event level to
represent the bound on the cross section.
Such stringent cuts reduce the $t\bar t$ background to less than or equal
to two events per year, whereas signal cross sections range from
66 (for case 3) to only 3 events (for case 6) per year.

As noted in Sec.~3, chargino pair production, followed by
the leptonic decays of charginos may well mimic this dilepton signal.
In order to get an idea of whether chargino production might be confused 
with slepton production, we generated chargino pair events with 
parameters as in cases 3 and 4 introduced above, and found that
no events passed the cuts (A-E). This leads to an upper limit of
0.1 fb [0.03 fb] for case 3 [case 4] so that at least for these
supergravity motivated parameter choices, the chargino signal is
unlikely to be confused with that of the slepton. The reason for this is,
of course, the very different kinematics in chargino and slepton events.
We expect that the chargino background in cases 5 and 6 will be even smaller.

The crucial cut for the detectability of sleptons over the background from
$t\bar t$ production is clearly
the central jet veto. The results presented above
assume 100\% jet detection efficiency, whereas real detectors
have regions of dead space. Without doing a real detector
simulation, we estimated the $t\bar t$ background assuming
a 99\% jet detection efficiency. In this case, some 
$t\bar t\to\ell\bar{\ell}+1$-jet events could be mistaken for 
$\ell\bar{\ell}+0$-jet events. We list the background for this imperfect
detector in column 7 of Table III, and see that in this case the background,
though still smaller than the signal, is certainly relevant in that it 
may considerably degrade the satistical
significance of any observed signal.

\subsection{Other Leptonic Signals}

Slepton production can also lead to other event topologies. The single
lepton signal, from $W^*\to\tell_L\tnu_{\ell}$
decays, is frequently larger than the dilepton signal discussed above.
There are, however, several SM sources which can lead to this event
topology. In Ref.\cite{AGUILA} it has already been noted that 
the background from $pp\to WZ\to l\nu +\nu\nu$ 
production is comparable to the slepton signal. The dominant background,
however, comes from single $W$ production. It might be expected that 
in this case, the transverse mass $M_T$($l$,$\eslt$) peaks sharply
at $M_W$, while the signal exhibits a broad distribution. Toward this
end, we have plotted, in Fig.~4, this distribution from the signal in 
cases 3 and 4,
and also from $W^*$, $WZ$ and $t \bar{t}$ production as given by the SM. We 
have required $\eslt \ge 100$ GeV, $p_T(\ell) > 20$ GeV, and $n(jets) = 0$. 
The transverse mass distribution from $W$ events do not exhibit the familiar
Jacobian peak because the hard $\eslt$ requirement we have imposed selects
out background mainly from off-shell $W$'s.
We see that over the entire range of $M_T$ where the signal is significant,
it is swamped by the single $W^*$ background. We thus conclude that detection
of the slepton signal in this channel would be extremely difficult.

Another possible strategy might be to examine the hadronically
quiet trilepton signal from
the cascade decays of the sleptons. We have done so for each of the four
LHC cases, and found no such events in our simulation in cases 5 and
6. In cases 3 and 4, we find signal cross sections of 4 fb and 1 fb,
respectively; 
however, the corresponding cross sections from $\tw_1\tz_2$ production
are 24~fb 
and 34~fb. We thus conclude that it is unlikely that trileptons from slepton 
sources will be
detectable above those from chargino-neutralino associated production even
if these signals are detectable above SM backgrounds\cite{BARBIERI}.

\section{Summary and Concluding Remarks}

We have studied the prospects for detecting sleptons of supersymmetry,
both at the Tevatron and at the proposed LHC. We find that the most
promising signal consists of events with acollinear leptons and no
jets. At the Tevatron, $W$ pair production is the dominant background,
while $t\bar{t}$ events, where the $b$ jets are too soft to be detected,
is the main background at the LHC. We have simulated the signal
as well as these backgrounds using ISAJET 7.03 with suitable cuts
to model the experimental conditions.

The signal cross sections we find are summarized in Tables I-III. We see
from Table I that even with an integrated luminosity of 1 fb$^{-1}$, the 
slepton signal is at most 50\% of the $WW$ background for the cases we
considered (the first of which has been picked to be nearly optimal within the 
constraints of supergravity models).
Incorporation of realistic detector effects
(such as cracks and imperfect electronics) would certainly reduce the
signal. We thus conclude that the detection of sleptons at the Tevatron
would be very difficult.

The situation is somewhat different for the LHC, where the higher energy
and the high design luminosity enables us to make strong cuts to enhance
the slepton signal over the background. We have designed two sets of cuts,
one each for sleptons lighter/heavier than about 200 GeV. The signal
cross sections after these cuts are shown in Table III together with
the corresponding cross sections from the $t\bar{t}$ and $WW$ backgrounds.
We find that sleptons lighter than 200 GeV should be
readily detectable with a year of LHC running at its design luminosity.
With the cuts designed for detection of heavier sleptons, we see that
the signal is extremely small (with no background in our simulation)
so that only a handful of events may be anticipated. We should also
warn the reader that this conclusion is based on the assumption that
the efficiency for jet detection is 100\%. If this falls by more than 1\%,
backgrounds from top quarks can become significant, so that the detectability
of heavy sleptons may be marginal. Finally, we have also examined
other channels for slepton detection. We find that the single lepton
channel is swamped by $W^*\to \ell\bar{\nu}$ events, 
whereas the rate for hadronically 
quiet
trilepton events coming from the cascade decays of sleptons is dwarfed by
that from direct $\tw_1\tz_2$ production. We also remark that we have
limited our attention to a strongly correlated set of input SUSY
parameters typical of supergravity models; breaking these correlations
could lead to a modification of our conclusions concerning slepton
detection ar hadron colliders.

To summarize, we are pessimistic about the prospects for detecting the sleptons
of supergravity at the Fermilab Tevatron. At the LHC, sleptons with
masses up to 200-250 GeV should be readily detectable with a year of running
at the design luminosity. Heavier sleptons might be also be detectable, but
this would crucially depend on the efficiency for vetoing events with 
central jets.

%
\acknowledgments

This research was supported in part by the U.~S. Department of Energy
under contract number DE-FG05-87ER40319, DE-AC35-89ER40486, and
DE-AM03-76SF00235. In addition, the work of HB was supported by the
TNRLC SSC Fellowship program. 
%
%
%
%

%
\newpage
%
%

\begin{table}
\caption[]{Cross sections in fb at Tevatron after cuts for cases~1 and
2 of slepton production, along with chargino production and SM
backgrounds.  We take $\mu =-m_{\tg}$ and $\tan\beta =2$.  Cuts are
described in the text.  The bound signifies the 1 event level in our
simulations.  Results have summed over $e$'s and $\mu$'s. }
\bigskip
\begin{tabular}{lrrrrrrr}
process & $m_{\tilde g}$ & $m_{\tell_L}$ & $m_{\tell_R}$ & $m_{\tnu}$ 
& $m_{\tw_1}$ & $m_{\tz_1}$ & $\sigma (fb)$ \\
\tableline
$\tell\bar{\tell}'(Case\ 1)$ & 150 & 88 & 80 & 62 & 63 & 24 & 18 \\
$\tell\bar{\tell}'(Case\ 2)$ & 200 & 112 & 102 & 93 & 73 & 32 & 9 \\
$\tw_1\overline{\tw}_1 (Case\ 1)$ & 150 & 88 & 80 & 62 & 63 & 24 & 5 \\
$\tw_1\overline{\tw}_1 (Case\ 2)$ & 200 & 112 & 102 & 93 & 73 & 32 & 3 \\
$WW$ & -- & -- & -- & -- & -- & -- & 36 \\
$t\bar t (150)$ & -- & -- & -- & -- & -- & -- & $<1$ \\
$Z\to\tau\bar\tau$ & -- & -- & -- & -- & -- & -- & $<9$ \\
\end{tabular}
\end{table}

\iftightenlines\else\newpage\fi

\begin{table}
\caption[]{Dilepton cross sections in fb and cut efficiency at LHC after cuts
for four cases of slepton production, along with backgrounds from top
quark and W-boson pair production.  We take
$\mu =-m_{\tg}$ and $\tan\beta =2$, and $m_t =150$~GeV.  
Cuts are described in the text.
Results have summed over $e$'s and $\mu$'s. }

\bigskip

\begin{tabular}{ccccccc}
cut & case 3 & case 4 & case 5 & case 6 & $WW$ & $t\bar{t}$ \\
\tableline
none & 3.5K & 243 & 55 & 17 & 100K & 1450K \\
A & 431(12\%) & 46(19\%) & 10(18\%) & 3.2(19\%) & 477(0.5\%) & 20K(1.4\%) \\
B & 62(14\%) & 22(48\%) & 6(60\%) & 2.5(78\%) & 11(2.3\%) & 3K(15\%) \\
C & 14(22\%) & 5(23\%) & 0.8(13\%) & 0.3(12\%) & 2(18\%) & 7.8(0.3\%) \\
D & 14(100\%) & 5(100\%) & 0.8(100\%) & 0.3(100\%) & 2(100\%) & 7.3(93\%) \\
\end{tabular}
\end{table}

\begin{table}
\caption[]{Cross sections in fb after final $p_T(\ell )$ and
$\Delta\phi (\ell\bar{\ell})$ cut along with background at LHC for
four cases of slepton production.  Parameters are as in Table 2.
Results have summed over $e$'s and $\mu$'s. }

\bigskip

\begin{tabular}{ccccccc}
case & $p_{T_c}$ (GeV) & $\Delta\phi_c$(deg) & signal & $t\bar{t}$(150) & 
$WW$ & $t\bar{t}$(1\%) \\
\tableline
case 3 & 40 & 90 & 2.2 & 0.06 & $<0.03$ & 0.33 \\
case 4 & 40 & 90 & 1.3 & 0.06 & $<0.03$ & 0.33 \\
case 4 & 80 & 140 & 0.8 & $<0.03$ & $<0.03$ & 0.09 \\
case 5 & 80 & 140 & 0.22 & $<0.03$ & $<0.03$ & 0.09 \\
case 6 & 80 & 140 & 0.1 & $<0.03$ & $<0.03$ & 0.09 \\
\end{tabular}
\end{table}


%
\begin{figure}
\caption[]{Slepton masses as a function of $m_{\tg}$ as given by Eq. 3,
for {\it a}) $m_{\tq}=m_{\tg}$ and {\it b}) $m_{\tq}=2m_{\tg}$. We plot
for $\tan\beta =2$ (solid) and 20 (dashes).}
\end{figure}
%
\begin{figure}
\caption[]{Total cross section for pair production of charged sleptons and sneutrinos
via DY mechanism versus slepton mass, for $p\bar{p}$ collisions at
$\sqrt{s}=1.8$~TeV.}The sleptons are taken to be degenerate.
\end{figure}
%
\begin{figure}
\caption[]{Total cross section for pair production of charged sleptons and sneutrinos
via DY mechanism versus slepton mass, for $p p$ collisions at
$\sqrt{s}=14$~TeV.}The sleptons are taken to be degenerate.
\end{figure}
%
\begin{figure}
\caption[]{Distribution in transverse mass for single lepton plus
no jet events from sleptons of case 3 and 4, and backgrounds from
$W^*\to \ell\nu_{\ell}$, $WZ$ and $t\bar t$ production, at
$\sqrt{s}=14$~TeV.}
\end{figure}
%
%

\vfill\eject

\centerline{\epsfbox{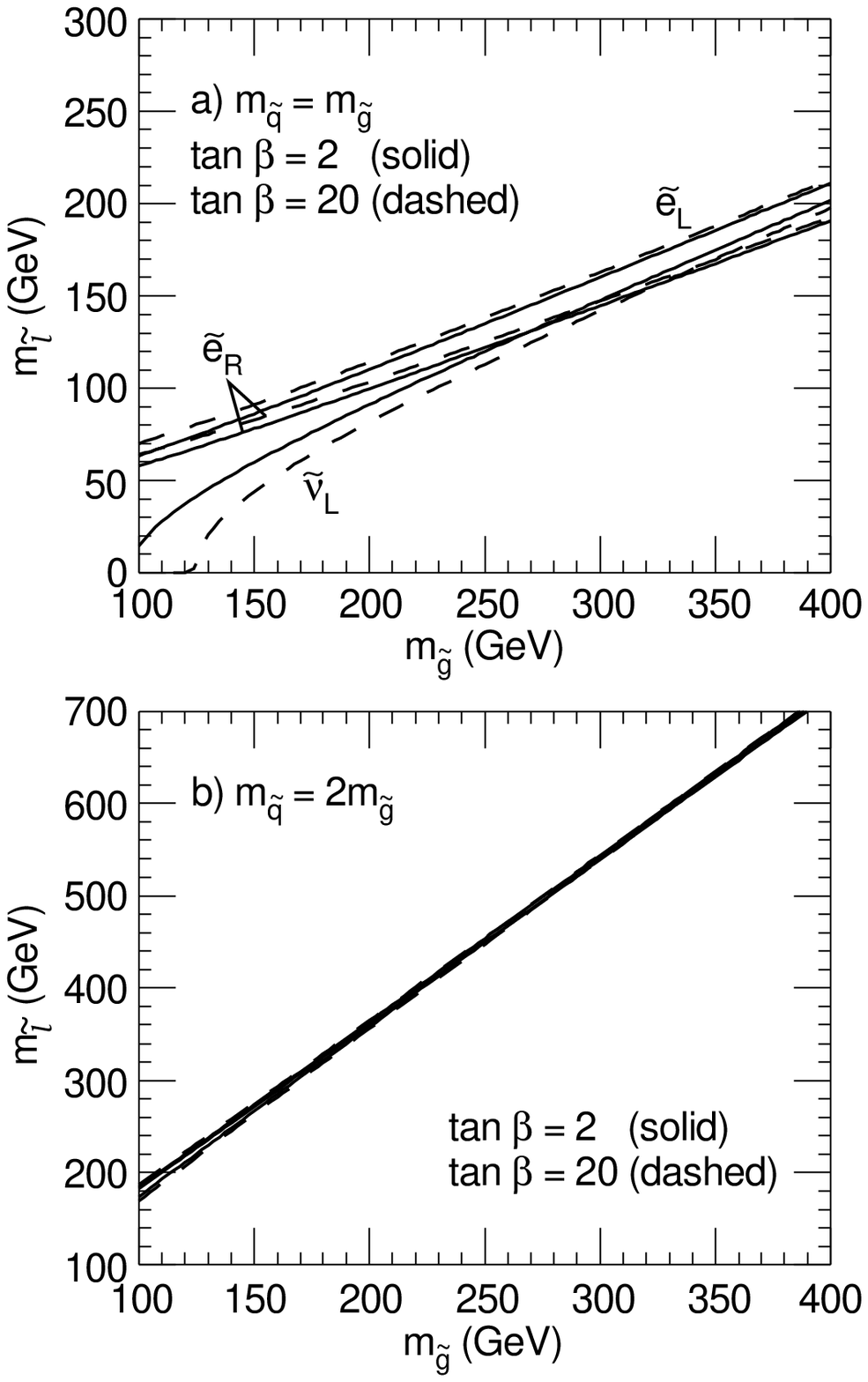}}
\bigskip\bigskip
\centerline{FIG.~1}
\vfill\eject

\centerline{\epsfbox{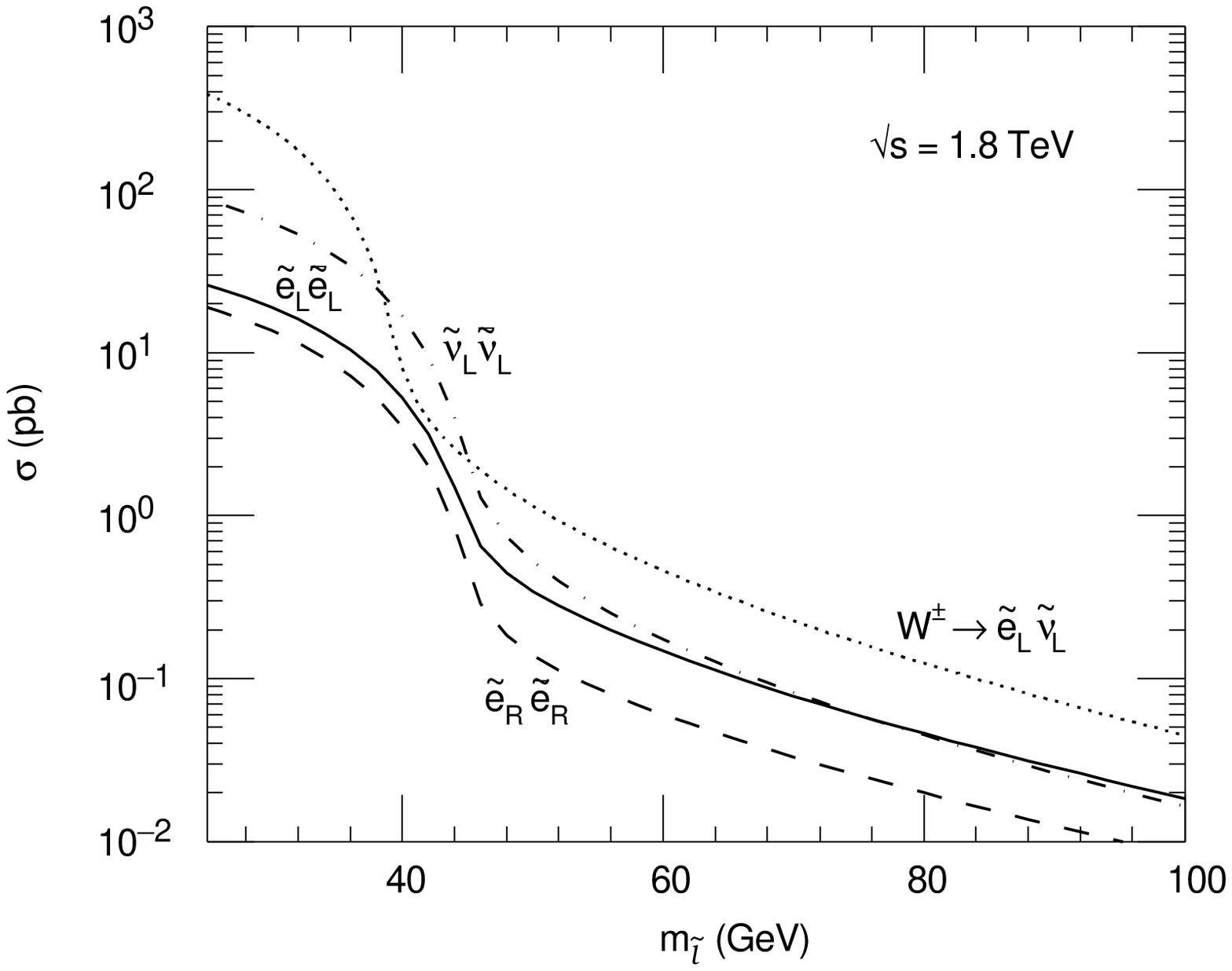}}
\bigskip\bigskip
\centerline{FIG.~2}
\vfill\eject

\centerline{\epsfbox{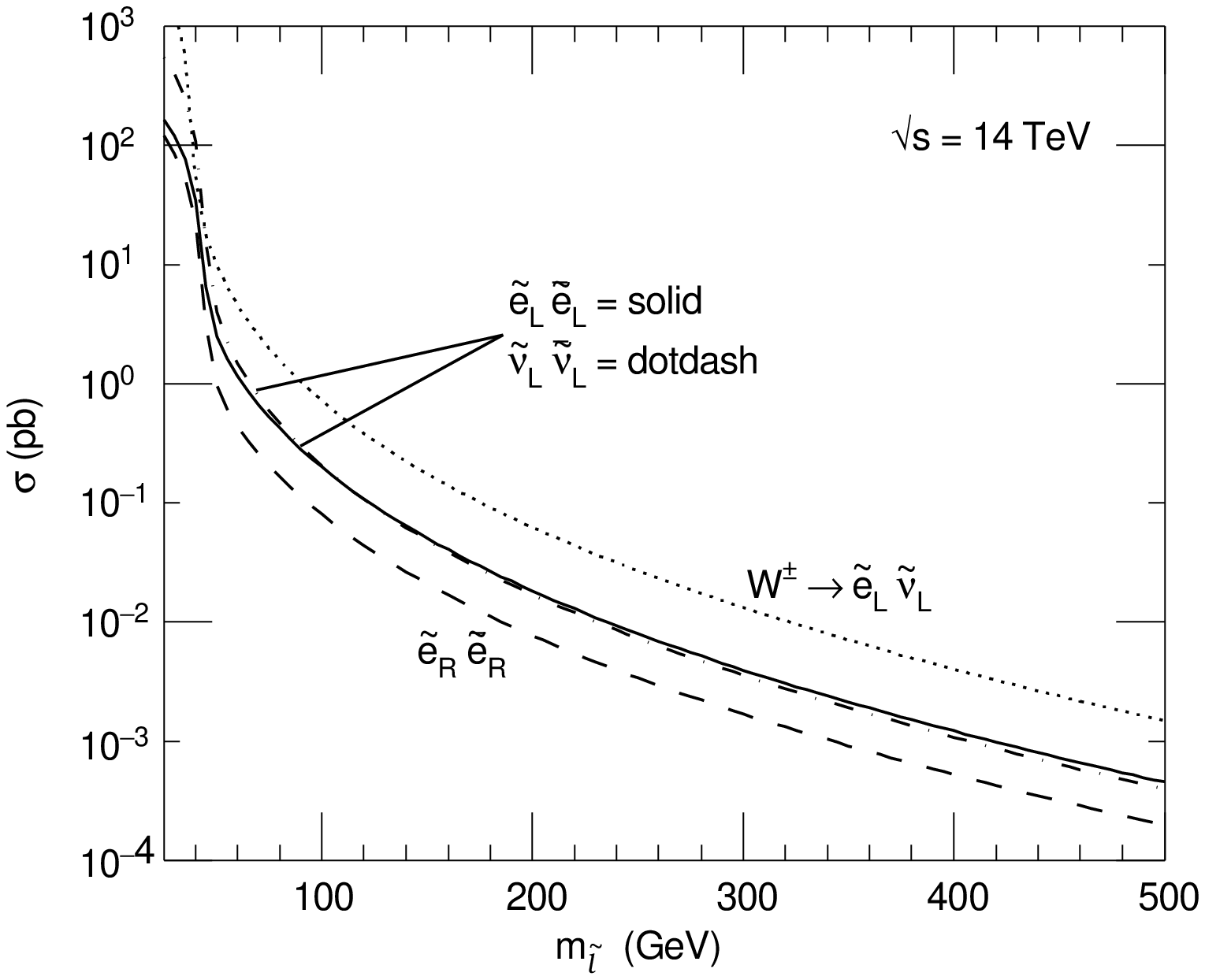}}
\bigskip\bigskip
\centerline{FIG.~3}
\vfill\eject

\centerline{\epsfbox{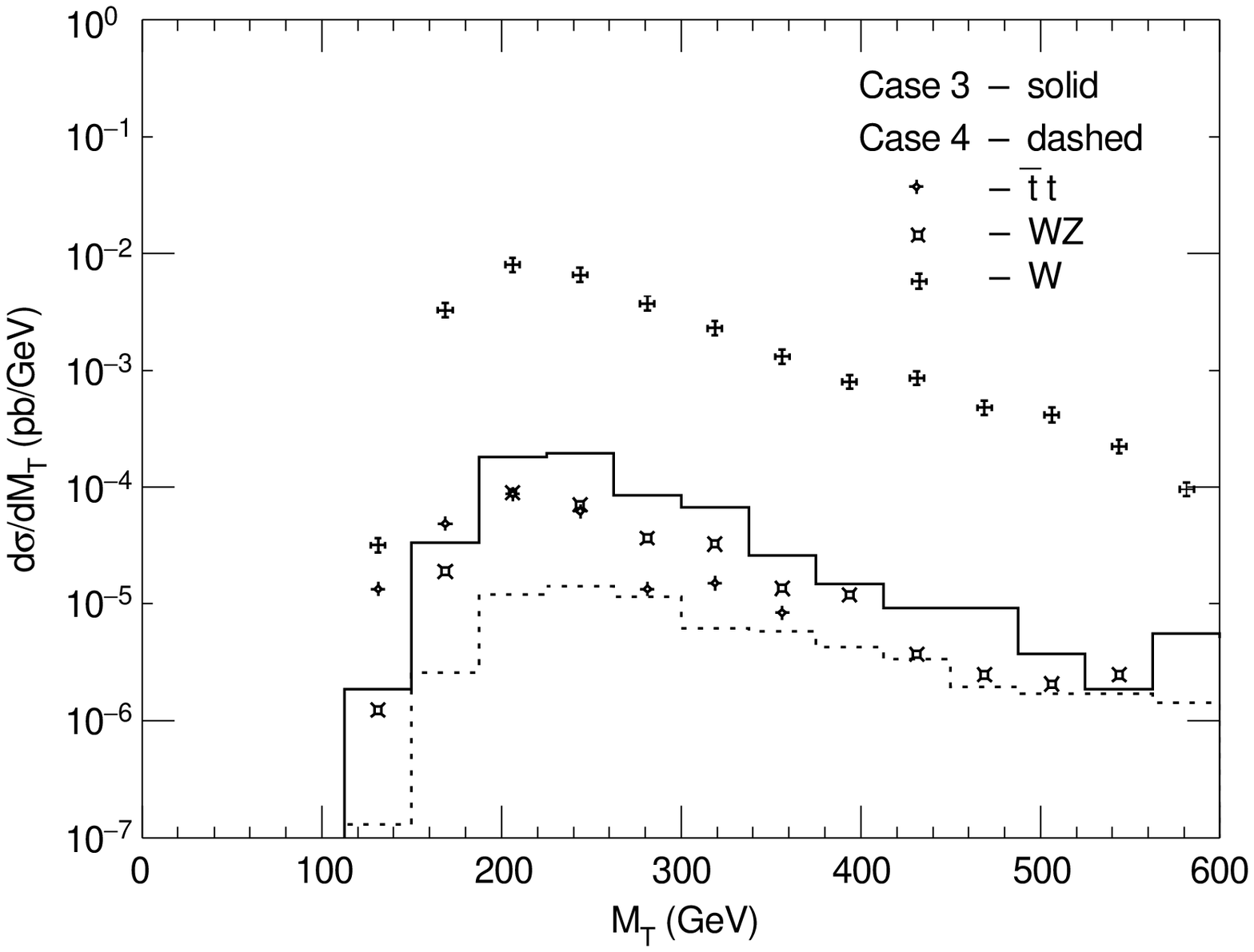}}
\bigskip\bigskip
\centerline{FIG.~4}
\vfill\eject

\end{document}